\documentclass[aps,pre,reprint,amssymb,superscriptaddress]{revtex4-1}
\usepackage{longtable}
\usepackage{graphicx}
\usepackage{braket}
\usepackage{amsmath}
\usepackage{upgreek}
\usepackage{here}
\usepackage{float}
\usepackage{sidecap}
\usepackage{color}

\begin{document}

\title{Expansion Dynamics After Laser-Induced Cavitation in Liquid Tin Microdroplets}

\author{Dmitry~Kurilovich}
\affiliation{Advanced Research Center for Nanolithography, Science Park~110, 1098~XG Amsterdam, The Netherlands}
\affiliation{Department of Physics and Astronomy, and LaserLaB, Vrije Universiteit Amsterdam, De Boelelaan 1081, 1081 HV Amsterdam, The Netherlands}

\author{Tiago~de~Faria~Pinto}
\affiliation{Advanced Research Center for Nanolithography, Science Park~110, 1098~XG Amsterdam, The Netherlands}
\affiliation{Department of Physics and Astronomy, and LaserLaB, Vrije Universiteit Amsterdam, De Boelelaan 1081, 1081 HV Amsterdam, The Netherlands}

\author{Francesco~Torretti}
\affiliation{Advanced Research Center for Nanolithography, Science Park~110, 1098~XG Amsterdam, The Netherlands}
\affiliation{Department of Physics and Astronomy, and LaserLaB, Vrije Universiteit Amsterdam, De Boelelaan 1081, 1081 HV Amsterdam, The Netherlands}

\author{Ruben~Schupp}
\affiliation{Advanced Research Center for Nanolithography, Science Park~110, 1098~XG Amsterdam, The Netherlands}

\author{Joris~Scheers}
\affiliation{Advanced Research Center for Nanolithography, Science Park~110, 1098~XG Amsterdam, The Netherlands}
\affiliation{Department of Physics and Astronomy, and LaserLaB, Vrije Universiteit Amsterdam, De Boelelaan 1081, 1081 HV Amsterdam, The Netherlands}

\author{Aneta~S.~Stodolna}
\affiliation{Advanced Research Center for Nanolithography, Science Park~110, 1098~XG Amsterdam, The Netherlands}
\affiliation{Department of Physics and Astronomy, and LaserLaB, Vrije Universiteit Amsterdam, De Boelelaan 1081, 1081 HV Amsterdam, The Netherlands}

\author{Hanneke~Gelderblom}
\affiliation{Department of Applied Physics, Eindhoven University of Technology, P.O. Box 513, 5600 MB Eindhoven, The Netherlands}

\author{Kjeld~S.~E.~Eikema}
\affiliation{Advanced Research Center for Nanolithography, Science Park~110, 1098~XG Amsterdam, The Netherlands}
\affiliation{Department of Physics and Astronomy, and LaserLaB, Vrije Universiteit Amsterdam, De Boelelaan 1081, 1081 HV Amsterdam, The Netherlands}

\author{Stefan~Witte}
\affiliation{Advanced Research Center for Nanolithography, Science Park~110, 1098~XG Amsterdam, The Netherlands}
\affiliation{Department of Physics and Astronomy, and LaserLaB, Vrije Universiteit Amsterdam, De Boelelaan 1081, 1081 HV Amsterdam, The Netherlands}

\author{Wim~Ubachs}
\affiliation{Advanced Research Center for Nanolithography, Science Park~110, 1098~XG Amsterdam, The Netherlands}
\affiliation{Department of Physics and Astronomy, and LaserLaB, Vrije Universiteit Amsterdam, De Boelelaan 1081, 1081 HV Amsterdam, The Netherlands}

\author{Ronnie~Hoekstra}
\affiliation{Advanced Research Center for Nanolithography, Science Park~110, 1098~XG Amsterdam, The Netherlands}
\affiliation{Zernike Institute for Advanced Materials, University of Groningen, Nijenborgh 4, 9747 AG Groningen, The Netherlands}

\author{Oscar~O.~Versolato}
\email{versolato@arcnl.nl}
\affiliation{Advanced Research Center for Nanolithography, Science Park~110, 1098~XG Amsterdam, The Netherlands} 

\date{\today}%

\begin{abstract}
The cavitation-driven expansion dynamics of liquid tin microdroplets is investigated, set in motion by the ablative impact of a 15-ps laser pulse. We combine high-resolution stroboscopic shadowgraphy with an intuitive fluid dynamic model that includes the onset of fragmentation, and find good agreement between model and experimental data for two different droplet sizes over a wide range of laser pulse energies. The dependence of the initial expansion velocity on these experimental parameters is heuristically captured in a single power law. Further, the obtained late-time mass distributions are shown to be governed by a single parameter. These studies are performed under conditions relevant for plasma light sources for extreme-ultraviolet nanolithography.
\end{abstract}

\maketitle

\section{Introduction}

Intense, short-pulse laser radiation can produce strong shock waves in liquids leading in some spectacular cases to explosive cavitation and violent spallation of the material \cite{stan2016liquid,stan2016negative,vinokhodov2016,basko2017,krivokorytov2017}. Such dramatic physical phenomena can readily find application, a very recent example of which is found in the field of nanolithography where microdroplets of liquid tin are used to create extreme ultraviolet (EUV) light~\cite{Banine2011,Pirati2017,Fomenkov2017}. These tin droplets, typically several 10\,$\mu$m in diameter, serve as targets for a high-energy, ns-pulse laser, creating a laser-produced plasma (LPP). Line emission from highly charged tin ions in the LPP provides the required EUV light. Currently, a dual-laser-pulse sequence is employed~\cite{Fomenkov2017}. In a first step, a ns-laser prepulse€ is used to carefully shape the droplet into a thin sheet that is considered to be advantageous for EUV production with a second, much more energetic, main pulse€. Recent developments \cite{Fomenkov2017, Kawasuji2017}, however, produced tentative but tantalizing evidence for significantly improved source performance when replacing the ns-prepulse with a ps-prepulse laser to produce a shock-wave-induced explosive fragmentation. Although some notable progress was made very recently~\cite{vinokhodov2016,basko2017,krivokorytov2017,Kriv2018}, this process requires further investigation.\\

\begin{figure}
\includegraphics[scale=1]{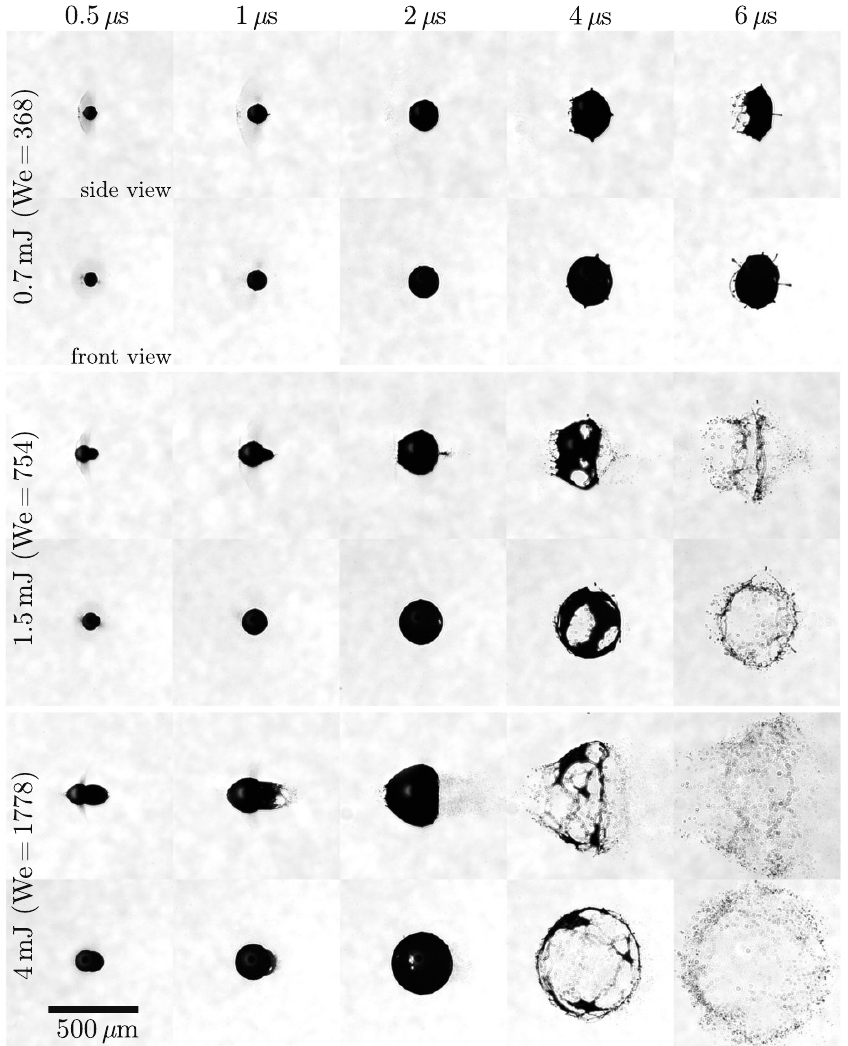}
\caption{\label{fig:summary} 
Stroboscopic shadowgraph images of expanding tin microdroplets (23\,$\mu$m initial radius) taken at different time delays for three different laser pulse energies (or Weber numbers, see Sec.~\ref{sec:Shell}) at a pulse length of 15\,ps, as seen from two viewing angles (90$^\circ$ side view and 30$^\circ$ front view). Laser impacts from the left; images are cropped and centered individually to improve visibility. A 500-$\mu$m-length scale bar is provided in the left-lower corner.}
\end{figure}

This paper advances the understanding of the aforementioned systems by providing an experimental and theoretical study of the late-time dynamics of the deformation of free-falling tin microdroplets. Being initially spherical, the droplets are subjected by strong shock waves generated by ps laser pulse impact, giving rise to cavitation \cite{vinokhodov2016,basko2017,krivokorytov2017,Kriv2018}. This centralized cavitation explosively propels the liquid to very high ($\sim$100\,m/s) radial velocities, producing a rapidly thinning liquid-tin shell (see Fig.\,\ref{fig:summary}). The initial spherical symmetry of the system is broken when the intensity of the shock wave exceeds a certain threshold and dramatic spallation is observed on the side of the droplet facing away from the laser impact zone. We focus our studies on the rich physics of the dynamics set in motion by the central cavitation.\\

We develop a model description for the time evolution of a stretching spherical shell, including the onset of fragmentation. Using stroboscopic shadowgraphic imaging, this model is experimentally validated over a wide range of laser pulse energies, 15\,ps in duration, and for two droplet sizes. The dependence of the initial expansion velocity on these experimental parameters is heuristically captured in a single power law. The late-time mass distributions, experimentally obtained from front-view shadowgraphy, are furthermore shown to be governed by a single parameter. 

\section{Experimental methods}
Our experimental setup has previously been described in detail \cite{Kurilovich2016}. For clarity, the most important characteristics, as well as experimental upgrades, are discussed here. A droplet generator is operated in a vacuum chamber ($10^{-7}$\,mbar) and held at constant temperature of 260$^{\circ}$\,C, well above the melting point of tin. The nozzle produces an on-axis multi-kHz train of 15 or 23\,$\mu$m radius droplets of 99.995\% purity tin, with density $\rho=7\,$g/cm$^3$ and surface tension $\sigma=0.532$\,N/m. The droplets relax to a spherical shape before they pass through a horizontal light sheet produced from a helium-neon laser. The light scattered by the droplets is detected by a photomultiplier tube, which signal is reduced in frequency to 10\,Hz to trigger a Nd:YAG laser system. This laser system produces tunable ps pulses of 1064\,nm wavelength light as described in more detail in Ref.\,\cite{Stodolna2018}. The laser pulse energy is varied between 0.5 and 5\,mJ employing a suitable combination of a waveplate and polarizer; its pulse length is kept constant at approximately 15\,ps for the experiments described here. The laser beam is focused down to a 100\,$\mu$m full-width-at-half-maximum (FWHM) diameter Gaussian spot. In order to maintain cylindrical symmetry, the laser light has circular polarization. An accurately timed laser pulse provides a radially-centered interaction with a falling droplet that occurs in a reproducible manner. Only a small fraction of the droplets interact with a laser pulse due to the mismatch in repetition rates of the droplet generator (multi-kHz) and laser~(10\,Hz).

Two shadowgraph imaging systems capture the dynamics of the expanding droplets. These systems are based on a single pulse from a broadband 560$\pm 2$\,nm wavelength 5-ns-pulse dye laser. Combined with two long-distance microscopes coupled with CCD cameras, this laser source provides back-lighting that is carefully rendered incoherent both temporally and spatially, with a 4\,$\mu$m resolution. One of these microscopes is aligned orthogonally to the laser beam to obtain side-view images; the other one is at 30$^\circ$ angle to the beam direction for a (tilted) front view. Both microscopes are equipped with bandpass filters to suppress the plasma radiation. The obtained images are used to track the size, shape, and velocity of the droplet expansion employing an image-processing algorithm.  Stroboscopic time series of different droplets are constructed by triggering the shadowgraphy systems once per drive laser shot with increasing delay, typically with 50--100\,ns time steps \cite{SM}.

\section{Results and interpretation}
The response of a tin droplet to laser pulse impact is shown in Fig.\,\ref{fig:summary} for three laser pulse energies. A qualitative description of the relevant physical processes leading to cavitation and spallation was recently given in Ref.~\cite{basko2017}. We summarize the crucial steps in the following, combining it with our experimental observations. The laser pulse impact ablates a thin (less than 1\,$\mu$m) layer of tin. At the lower pulse energies (see the 0.7 or 1.5\,mJ cases presented in Fig.\,\ref{fig:summary}), this ablated mass is clearly visible and seen to move away from the droplet in the direction opposite to the laser light. The remarkably sharp outer boundary of this ablated mass may be explained by the existence of an inhomogeneous two-phase, gas-liquid mixture of low average density but approaching liquid density in the vicinity of the ablation front \cite{Tinten1998,VONDERLINDE2000}. In contrast to the ns laser pulse impact \cite{Kurilovich2016}, the resulting droplet propulsion is limited here.

The ablation pulse gives rise to a pressure wave starting from the laser-impacted region of the droplet. This pressure is applied on a short, ps time scale that is several orders of magnitude shorter than the time scale on which pressure waves travel through the droplet.

As a consequence, the liquid is locally compressed and pressure waves travel through the droplet \cite{reijers2017}. Converging pressure "shock" waves superimpose in the center of the droplet where they cause tensile stress. Once this tensile stress is above the yield strength of the liquid it ruptures and a thin shell is formed. Theoretical modeling \cite{basko2017} has shown that there is a high sensitivity of both the rate of expansion and the morphology of the deformed droplet to the details of the, poorly known, metastable equation-of-state (EOS) of tin in the region of the liquid-vapor phase transition and to the parameters of the critical point. Thus, precise theory predictions of the post-cavitation dynamics of liquid tin microdroplets cannot yet be made and, instead, measurements as presented in this work may provide a sensitive instrument for probing the EOS of liquid metals \cite{basko2017}.

After passing through the droplet center, the shock wave reaches the back side of the droplet (i.e. the side facing away from the laser) and reflects from this free interface where it can again rupture the liquid and give rise to spallation that rapidly propels, at several $\,100\,$m/s, a small mass fraction of the droplet along the laser beam propagation direction (see Fig.\,\ref{fig:summary}) \cite{basko2017,krivokorytov2017}. The fluid dynamic description of this spallation is very rich and will be left for future work. However, it is noteworthy that there is a regime where a strong spallation occurs, but is mitigated under capillary action (compare the $t$\,=\,1 and 2\,$\mu$s shadowgraphs for the 1.5\,mJ case in Fig.\,\ref{fig:summary}).

We note that the higher-energy cases in Fig.\,\ref{fig:summary} also hint at the existence of a hole on the side of the droplet facing the laser, which may possibly be attributed to ablation-thinning of the droplet shell. In combination with the spallation, a "tunnel" is thus created (see the front view shadowgraphs in Fig.\,\ref{fig:summary}), that may later form a doughnut-type mass distribution. Another scenario, when fragmentation following the collapse of the shell results in high-speed jetting, corresponding to the lowest laser pulse energies, is reported on in Refs.~\cite{thoroddsen2009, gonzalez2011, Kriv2018}.\\ 

In the following, we focus on the late-time dynamics set in motion by the central cavitation. Firstly, a basic model for the time evolution of the intact liquid shell is presented. Secondly, we study the time scale at which first holes in this shell become visible, after which full fragmentation of the shell rapidly sets in. This results, eventually, in a late-time mass distribution, the understanding of which is of particular relevance for producing EUV radiation at high conversion efficiencies.\\

\subsection{\label{sec:Shell}Time evolution of the shell}
\begin{figure}[]
\includegraphics[scale=1]{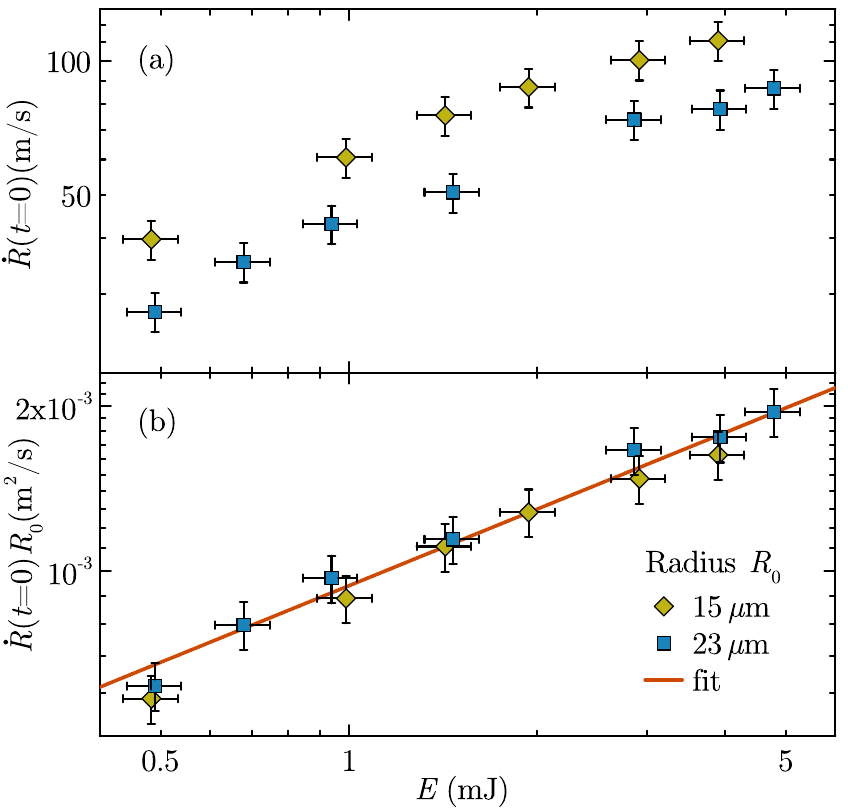}
\caption{\label{fig:Rdot}(a) Initial radial expansion velocity $\dot{R}$($t$=0) of tin microdroplets as a function of total laser pulse energy~$E$. Yellow diamonds are for $R_0$\,=\,15\,$\mu$m and blue squares are for $R_0$\,=\,23\,$\mu$m. The error bars represent 10$\%$  uncertainty on the velocity measurements and 20$\%$ uncertainty on the energy measurements. (b) The same data, rescaled by multiplying by the initial droplet radius~$R_0$. The solid red line is obtained by fitting a power-law function to the concatenated data.}
\end{figure}

The shadowgraphs such as those presented in Fig.\,\ref{fig:summary} clearly show a cylindrically symmetric expanding shell, the radius $R(t)$ of which is tracked for each time step by measuring the maximal transverse (vertical in Fig.\,\ref{fig:summary}) size of this feature from the side-view shadowgraphs. 
By fitting a linear function to the first few $R(t)$ data points after the laser impact, we obtain values for the initial droplet expansion velocity $\dot{R}$($t$=0). Figure~\ref{fig:Rdot}(a) shows a monotonic increase of this radial expansion velocity $\dot{R}$($t$=0) as a function of laser pulse energy. Using a heuristic argument, we collapse all obtained data onto a single curve, scaling the obtained velocities with the initial droplet radius $R_0$ (see Fig.\,\ref{fig:Rdot}(b)). To arrive at this scaling, we hypothesize that the magnitude of the induced shock wave, and its related cavitation event, scales with the laser-facing surface area, $\propto$\,$R_0^{2}$ (the droplet is much smaller than the laser spot size). Meanwhile, the expansion is impeded by the droplet mass $\propto$\,$R_0^{-3}$. By neglecting the possible differences in the dissipation of the shock wave, one thus arrives at the aforementioned simple scaling  $\dot{R}$($t$=0$)$\,$\propto$\,$R_0^{-1}$. As seen from Fig.\,\ref{fig:Rdot}(b), such a rescaling causes all data to fall on a single curve that appears to be represented quite well by a power-law function $\dot{R}$($t$=0$)R_0$\,=\,$A$\,$\times$\,$E^\alpha$ with fit parameters $A$\,=\,9.4$\pm1.4$\,$\rm{m^2}/s$ and $\alpha$\,=\,0.46$\pm0.02$. Although such power-law behavior is well established in the relevant literature on ablation \cite{Eidmann1984,Dahmani1993,Eidmann2000}, these fit parameters cannot be straightforwardly predicted from available theory, given their sensitivity to the EOS and the details of laser-matter interaction.\\

Next, to describe the time evolution of the liquid droplets in Fig.\,\ref{fig:summary}, we hypothesize that a small ($\sim$$\mu$m), high-pressure ($\sim$kbar) cavitation bubble expands the liquid quickly into a thin shell. Thus, a spherically symmetric shell expands at a certain initial velocity $\dot{R}$($t$=0), with the cavitation pressure having done its thermodynamic work effectively at time zero. Given that the droplet expands into the vacuum, the only limitation to its expansion is surface tension.

The liquid tin shell has thickness $h(t)$ which decreases quadratically with time as mass conservation implies, in the limit of a thin shell ($R(t)$\,$\gg$\,$h(t)$), that $h(t)=R_0^3/(3R(t)^2)\approx R_0^3/(3t^2\dot{R}(t$=0)$^2)$. The expansion of the shell can be described by the Rayleigh-Plesset equation, which in the limit of a thin shell reads \cite{Vledouts2016}: $p(R-h)-p(R)\approx \rho h \ddot{R}$, where $p$ is pressure. 
 Imposing dynamic boundary conditions at the two liquid-vapor interfaces $p(R-h)$ and $p(R)$, and noting that the only force acting on the shell is the Laplace pressure, the net value of which is given by $p=4\sigma/R(t)$ (in contrast to Ref.~\cite{Vledouts2016}, where this contribution was negligible), we find 
\begin{equation}
\ddot{R}(t)=\frac{-p}{\rho h(t)}=-\frac{4\sigma}{\rho h(t)  R(t) }=-\frac{12\sigma R(t)}{\rho R_0^3}. 
\end{equation}
From the above expression, using the boundary conditions of a given $\dot{R}(t$=0) and initial radius $R_0$, one obtains
\begin{equation}
\frac{R(t)}{R_0}=\cos(\pi t/\tau_c)+\sqrt{\frac{\textrm{We}}{12}}\sin(\pi t/\tau_c),\label{eq:R(t)}
\end{equation}
introducing the capillary time scale here as $\tau_c$\,=\,$\pi \sqrt{\rho R_0^3/12\sigma}$ and the relevant Weber number as We\,=\,$\rho R_0\dot{R}$($t$=0$)^2/\sigma$. The above solution is identical in form to that obtained previously for a disk-type expansion of a droplet following ns-laser pulse impact~\cite{Gelderblom:2015,Kurilovich2016}. 

\begin{figure}[]
\includegraphics[scale=1]{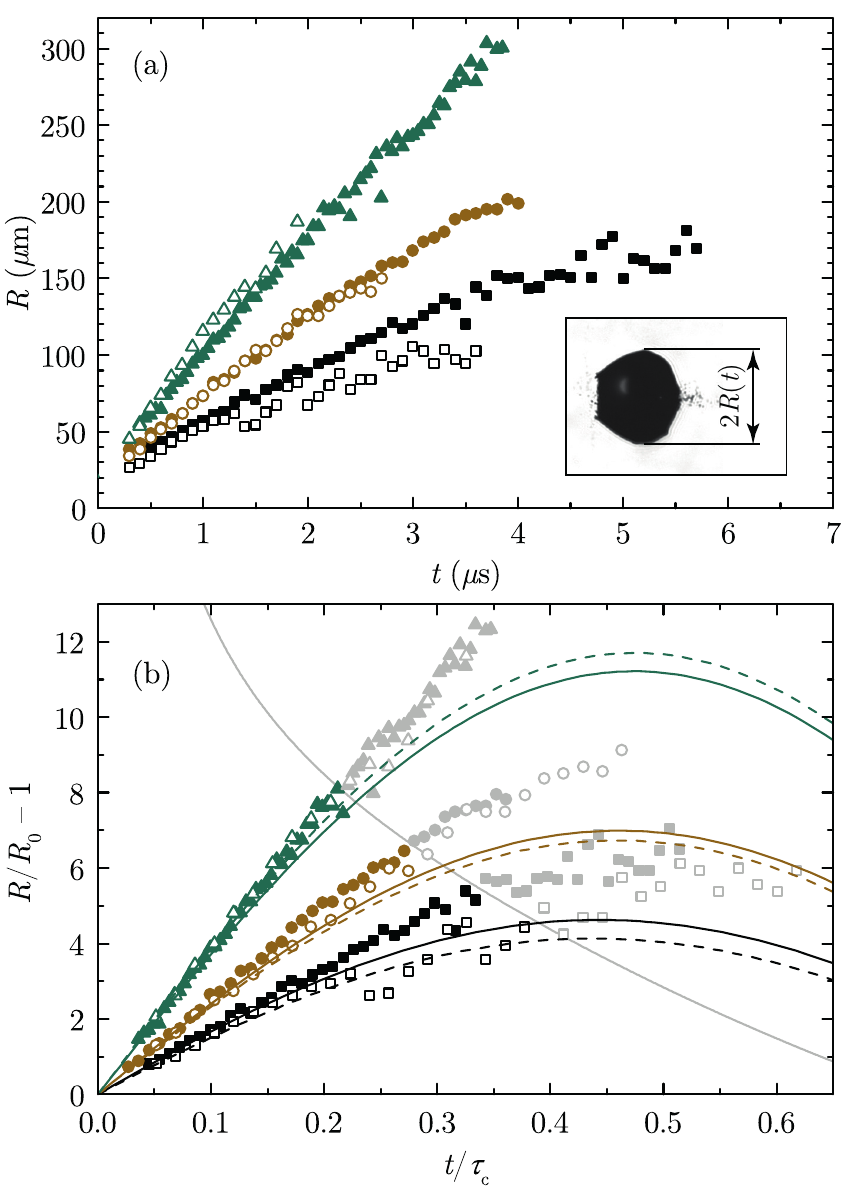}
\caption{\label{fig:dynamics}(a) Radius $R(t)$ of shell driven by cavitation in $R_0=[15]\,23\,\mu$m droplets for similar Weber numbers. [Open] black squares: $\rm{We}=[304]\,368$; [open] brown circles: $\rm{We}=[705]$\,754; [open] green triangles $\rm{We}=[1923]\,1778$. The inset (500\,$\mu$m\,$\times$\,700\,$\mu$m) represents a side-view image of the $\rm{We}=754$ case at 2.5\,$\mu$s. (b) Rescaled, dimensionless radius as a function of dimensionless time $t/\tau_c$. The solid curves represent the model predictions (see Eq.\,(2)) for the 23\,$\mu$m initial radius droplet; the dashed curves show the same but for the smaller, 15\,$\mu$m droplet. The solid gray line depicts the destabilization radius $R(t_*)$ as obtained from the fit of Eq.\,(\ref{eq:tstar}) to the data (see Fig.\,\ref{fig:holedata}), applied to Eq.\,(\ref{eq:R(t)}). The data are grayed out beyond this line.}
\end{figure}

Next, we apply the theory developed above with zero fit parameters taking as an input the experimental values for $\dot{R}(t$=0). The results are presented in Fig.\,\ref{fig:dynamics}(b), where the experimentally obtained radius $R(t)$ (see Fig.\,\ref{fig:dynamics}a) is rendered dimensionless by dividing by $R_0$ (and subtracting the unity value offset), and plotted as a function of dimensionless time $t/\tau_c$. Given the simplicity of our arguments, we find excellent agreement between model and experiment. The observed late-time "overshoot" that is apparent in the experimental data shown, with theory predicting a more rapid retraction of the shell, can be understood by considering the process of hole formation and ensuing fragmentation, which strongly reduces the restoring forces. \\

\subsection{Hole opening time}

\begin{figure}[t]
\includegraphics{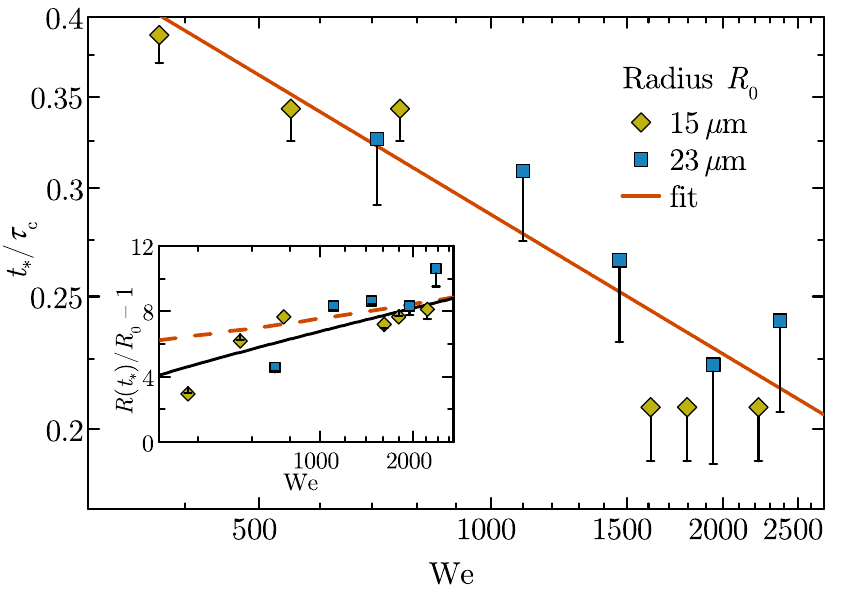}
\caption{\label{fig:holedata} Dimensionless time of first hole opening $t_*/\tau_c$ as a function of Weber number for two droplet sizes: 15\,$\mu$m (yellow diamonds) and 23\,$\mu$m initial radius (blue squares). For We\,$\lesssim$\,300, the shell collapsed before holes were apparent. Error bars are two time steps towards shorter values. The solid red line represents the best fit of a power law to the concatenated data. The inset graph depicts the dimensionless radius of the shell at the moment of hole opening. The solid black line shows the destabilization radius as obtained from inserting Eq.\,(\ref{eq:tstar}) into Eq.(\,\ref{eq:R(t)}). The dashed red line is based on Eq.\,(\ref{eq:Rtstar}) taking $\beta$\,=\,0.34 (see main text).}
\end{figure}

The expanding shell is subjected to Rayleigh-Taylor instabilities (RTI), here as instabilities of the liquid-vapor interface driven by the radial acceleration and deceleration of the liquid sheet \cite{Vledouts2016}. Surface tension provides mode selection of the RTI \cite{Keller1954,Chandrasekhar1961,bremond2005bursting,Vledouts2016}. This phenomenon will lead to hole formation after which rapid hole opening and merging will lead to full fragmentation of the shell. In our system, there are two acceleration mechanisms that act along the surface normal and that thus will contribute to RTI growth. Firstly, the liquid is strongly accelerated by the cavitation pressure. Secondly, after the initial fast expansion, the shell much more slowly decelerates under the influence of surface tension. Holes will form at the shell piercing time $t_*$, when the size of a growing instability is of the order of the shell thickness~\cite{Vledouts2016}.
In the available literature (e.g., see Refs.\,\cite{Vledouts2016,bremond2005bursting}), quite generally a scaling relation is found of the type
\begin{equation}
\frac{t_*}{\tau_c} \propto \textrm{We}^{-\beta}, \label{eq:tstar}
\end{equation}
resulting in a rough scaling for the corresponding destabilization radius \cite{Vledouts2016} 
\begin{equation}
\frac{R(t_*)}{R_0} - 1\propto \textrm{We}^{1/2-\beta}.\label{eq:Rtstar}
\end{equation}
Here we have, for simplicity, linearized the expansion rate to a constant $\dot{R}$\,$\propto$\,We$^{1/2}$ and in Eq.\,(\ref{eq:tstar}) dropped the usual weak scaling term $(\eta_0/R_0)^{\beta}$,  where $\eta_0$ is the initial amplitude of RTI. The positive power $\beta$ is typically smaller than unity \cite{bremond2005bursting,Vledouts2016,KleinPhDThesis}. Obtaining a theory value for this power is complicated because of the two competing mechanisms driving RTI. We take instead the rather general form of Eq.\,(\ref{eq:tstar}) and let the experiment provide the relevant value for $\beta$ and for the proportionality constant. In the experiments, a hole becomes clearly visible in side view shadowgraphy only when another hole simultaneously appears on the opposite side. This naturally introduces a slight detection bias. However, it is possible to reliably and reproducibly obtain an estimate for $t_*$ based on optical inspection of the experimental data supported by the fact that for times $t$\,$>$\,$t_*$ (where we take 100\,ns time steps) the shell becomes permeated with holes. We observe a monotonically decreasing value for $t_*$ with increasing Weber number as expected from Eq.\,(\ref{eq:tstar}). Furthermore, $t_*$ is seen to decrease with droplet size for similar Weber number.
Rescaling $t_*\to t_*/\tau_c$ in Fig.\,\ref{fig:holedata} indeed collapses the data for the two different droplet sizes onto a single curve. The resulting fit values of the power law Eq.\,(\ref{eq:tstar}) are given by $\beta$\,=\,0.34(4) and 3.0(8) for the proportionality constant. As we clearly find $\beta$\,$<$\,1/2 we can now immediately conclude from Eq.\,(\ref{eq:Rtstar}) that the destabilization radius $R(t_*)$ increases with Weber number, and thus with laser energy. The obtained fit values will serve as input for predicting the late-time mass distribution produced by the laser impact.\\

\begin{figure}[b]
\includegraphics[scale=1]{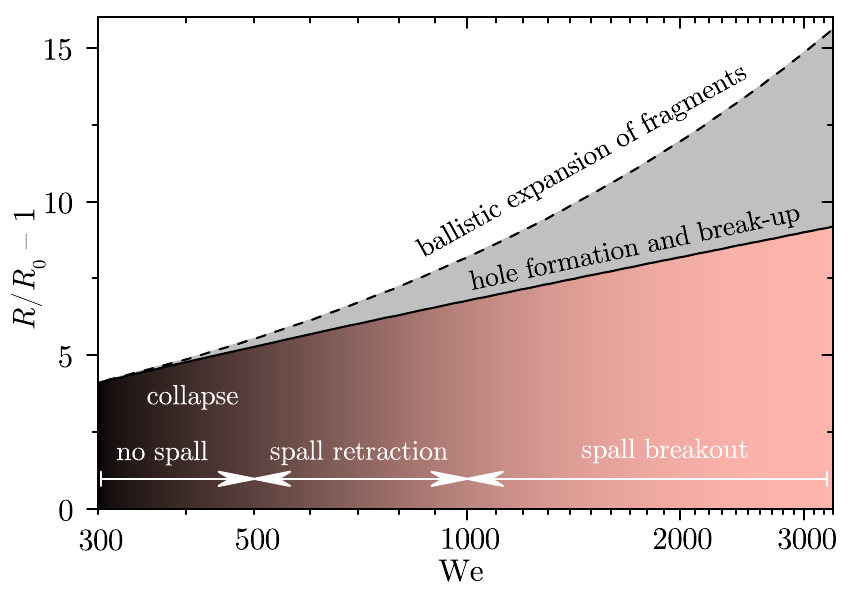}
\caption{\label{fig:master} Phase diagram depicting the maximum obtainable radius as a function of Weber number. White arrows illustrate the various spallation regimes (see main text). The dashed line is obtained by maximizing Eq.\,(\ref{eq:R(t)}). The solid line shows the destabilization radius as obtained from inserting Eq.\,(\ref{eq:tstar}) into Eq.(\,\ref{eq:R(t)}); the area under this line is obtainable without holes occurring.}
\end{figure}

\subsection{Late-time mass distribution}
Having achieved good agreement between experiment and model, we now summarize our findings in Fig.\,\ref{fig:master} to facilitate a more direct, industrial application of our findings. The maximum amplitude of an intact retracting shell is given by maximizing Eq.\,(2). The resulting dashed black curve in Fig.\,\ref{fig:master} is close to the well-known scaling $\sim$$\sqrt{\textrm{We}}$ (as was pointed out in Ref.\,\cite{Klein2015}). The maximum obtainable shell size without any holes (i.e. the destabilization radius) is obtained by inserting Eq.\,(\ref{eq:tstar}) into Eq.\,(\ref{eq:R(t)}) as in Fig.\,\ref{fig:dynamics}. The result is shown as a solid black line in Fig.\,\ref{fig:master}. Once this line is crossed in the expansion phase, the rapid breakup and fragmentation of the shell, associated with large Weber numbers here, precludes capillary collapse and the droplet target fragments ballistically expand. For the low Weber numbers shown the shell collapses and jetting may ensue \cite{Kriv2018}.\\

In Fig.\,\ref{fig:master} we have also indicated ranges of Weber numbers that we associate with different spallation regimes. For Weber numbers below We\,$\approx$\,500, no spallation is visible at all; for 500\,$\lesssim$\,We\,$\lesssim$\,1000 the spall is nearly fully retracted due to surface tension; and for We\,$\gtrsim$\,1000 there is a rather abrupt and violent breakout of the spall. For We\,$\gtrsim$\,1000, we find that hole formation in the "tunnel" wall leads very quickly to complete breakup of the shell, where the fragments continue on their ballistic trajectories with radial velocities close to the initial $\dot{R}(t$=0).\\

\begin{figure}[t]
\includegraphics[scale=1]{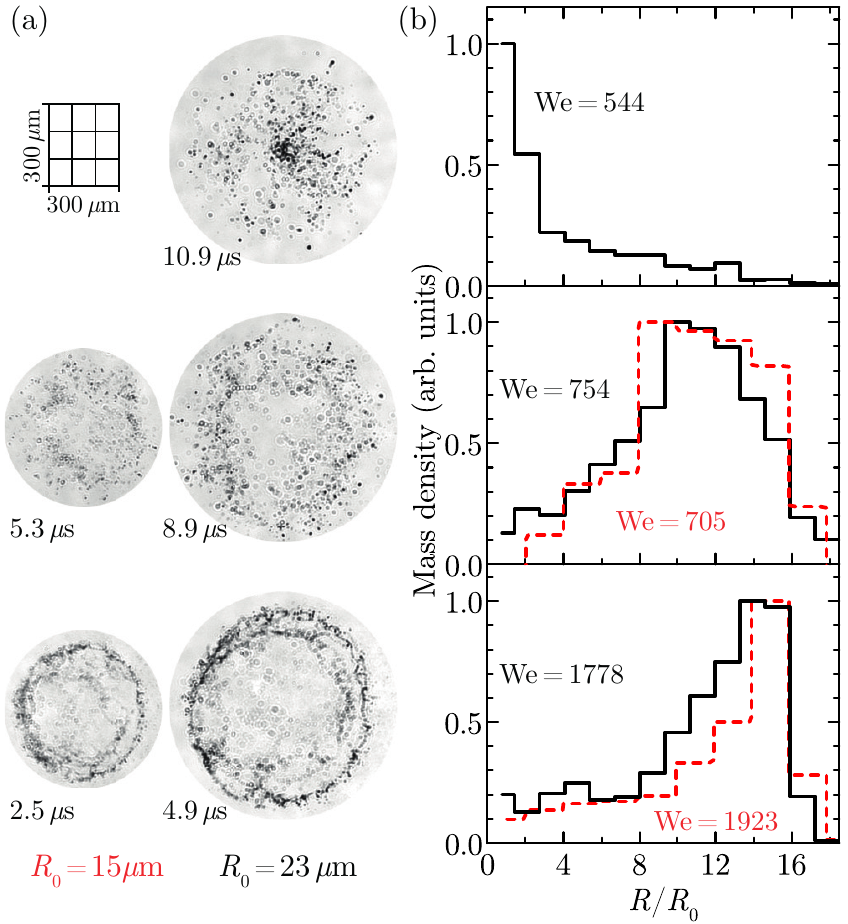}
\caption{\label{fig:finalmass} (a) Late-time front (30$^\circ$) view shadowgraphs of expanded and deformed droplets of different droplet sizes representing cases of similar Weber numbers. The rectangular scale grid in the top-left corner visualizes the effect of the $30^\circ$ parallax angle. (b) Inverted, angularly averaged, radial projections of the obtained gray scale of corresponding images from (a), which gives a qualitative measure of the radial column-density-mass distribution. The dashed red line corresponds to the data from $R_0$\,=\,15\,$\mu$m droplets; the solid black line is for $R_0$\,=\,23\,$\mu$m.}
\end{figure}

From an application perspective, it is interesting to study the "final" late-time fragment mass distribution that might serve as a target for a main laser pulse in an actual industrial EUV-source based on tin plasma. In the following analysis, we present some brief guidelines based on the here presented data but keep the full analysis of the relevant fragmentation process (including fragment size distributions) for a later, dedicated work. We find that the final mass distribution changes very dramatically over a relatively small range of Weber numbers as can be seen from the shadowgraphs presented in Fig.\,\ref{fig:finalmass}(a) as well as from their angularly averaged, radial projections shown in Fig.\,\ref{fig:finalmass}(b). These projections can loosely be interpreted as a column-density mass distribution along the line-of-sight of the backlight illumination. The projections are not corrected for the small parallax angle or for the limited depth of focus. Still, we expect to track a large fraction of the total droplet mass. Similar mass distributions are found for 15 and 23\,$\mu$m droplets at comparable Weber numbers (see Fig.\,\ref{fig:finalmass}).

For We\,=\,544 we observe that the mass distribution has a maximum in the center of the image, as is to be expected from a collapse. For just very slightly larger Weber numbers, we find instead a toroidal profile that becomes more and more pronounced for larger Weber numbers. We attribute this observation to the RTI-driven breakup of the "tunnel" wall (thus preventing collapse) as well as by a pronounced spallation that modifies the mass distribution, as was also noted in Ref.\,\cite{basko2017}.

The here established and quantified sensitivity of the final state mass distribution to the Weber number can readily find application. As an example, we note that a typical target size used in the industry is in the order of a few hundred micrometers \cite{Fomenkov2017}, several $\mu$s after prepulse impact. In our experiments, a relevant $\sim$450\,$\mu$m diameter target is, for example, obtained from a $R_0$\,=\,15\,$\mu$m initial droplet size after expanding for 2.5\,$\mu$s, see Fig.\,\ref{fig:finalmass}(a) for We\,=\,1923. In this case, the main pulse would not find a significant mass fraction in its focus. Thus, it may well be opportune to choose a lower Weber number, by tuning down the laser energy, and obtain a more homogeneous mass distribution which could improve the conversion efficiency of drive laser light into useful EUV radiation by, e.g., improving the absorption of drive laser light.\\

\section{Conclusions}
We present an analysis of the cavitation-driven expansion dynamics of liquid tin microdroplets that is set in motion by the ablative impact of a 15-ps laser pulse. High-resolution stroboscopic shadowgraphy of the expanding tin shells is combined with an intuitive fluid dynamic model that includes the time and size at which the onset of fragmentation becomes apparent. This model will aid follow-up studies of the fragmentation pathways. Good agreement between model and experimental data is found for two different droplet sizes over a wide range of laser pulse energies. The dependence of the initial expansion velocity of the liquid shell on these experimental parameters is heuristically captured, for application purposes, in a single power law. A summary phase diagram of the expansion dynamics is presented. It covers regimes with and without spallation, as well as a transition regime where the spalled material is retracted under surface tension. This transition regime enables finding conditions for a maximum shell expansion velocity with strongly suppressed forward-moving debris. This phase diagram facilitates a more direct, industrial application of our findings. 

Further, the experimentally obtained late-time mass distributions are shown to be governed by a single parameter, the Weber number. These studies are of particular relevance for plasma sources of extreme ultraviolet light for nanolithography. In such plasma sources, the tin mass distributions obtained upon cavitation-driven shell fragmentation, as studied in this work, are shown to be promising targets for efficient laser coupling \cite{Fomenkov2017, Kawasuji2017}.\\

\section*{Acknowledgments}
We thank A.~L.~Klein and M.~M.~Basko for insightful discussions. This work has been carried out at the Advanced Research Center for Nanolithography (\mbox{ARCNL}), a public-private partnership between the University of Amsterdam, the Vrije Universiteit Amsterdam, the Netherlands Organization for Scientific Research (NWO) and the semiconductor equipment manufacturer ASML.\\
\clearpage
\bibliographystyle{apsrev4-1}
\bibliography{bib}

\begin{thebibliography}{28}%
\makeatletter
\providecommand \@ifxundefined [1]{%
 \@ifx{#1\undefined}
}%
\providecommand \@ifnum [1]{%
 \ifnum #1\expandafter \@firstoftwo
 \else \expandafter \@secondoftwo
 \fi
}%
\providecommand \@ifx [1]{%
 \ifx #1\expandafter \@firstoftwo
 \else \expandafter \@secondoftwo
 \fi
}%
\providecommand \natexlab [1]{#1}%
\providecommand \enquote  [1]{``#1''}%
\providecommand \bibnamefont  [1]{#1}%
\providecommand \bibfnamefont [1]{#1}%
\providecommand \citenamefont [1]{#1}%
\providecommand \href@noop [0]{\@secondoftwo}%
\providecommand \href [0]{\begingroup \@sanitize@url \@href}%
\providecommand \@href[1]{\@@startlink{#1}\@@href}%
\providecommand \@@href[1]{\endgroup#1\@@endlink}%
\providecommand \@sanitize@url [0]{\catcode `\\12\catcode `\$12\catcode
  `\&12\catcode `\#12\catcode `\^12\catcode `\_12\catcode `\%12\relax}%
\providecommand \@@startlink[1]{}%
\providecommand \@@endlink[0]{}%
\providecommand \url  [0]{\begingroup\@sanitize@url \@url }%
\providecommand \@url [1]{\endgroup\@href {#1}{\urlprefix }}%
\providecommand \urlprefix  [0]{URL }%
\providecommand \Eprint [0]{\href }%
\providecommand \doibase [0]{http://dx.doi.org/}%
\providecommand \selectlanguage [0]{\@gobble}%
\providecommand \bibinfo  [0]{\@secondoftwo}%
\providecommand \bibfield  [0]{\@secondoftwo}%
\providecommand \translation [1]{[#1]}%
\providecommand \BibitemOpen [0]{}%
\providecommand \bibitemStop [0]{}%
\providecommand \bibitemNoStop [0]{.\EOS\space}%
\providecommand \EOS [0]{\spacefactor3000\relax}%
\providecommand \BibitemShut  [1]{\csname bibitem#1\endcsname}%
\let\auto@bib@innerbib\@empty
\bibitem [{\citenamefont {Stan}\ \emph
  {et~al.}(2016{\natexlab{a}})\citenamefont {Stan}, \citenamefont
  {Milathianaki}, \citenamefont {Laksmono}, \citenamefont {Sierra},
  \citenamefont {McQueen}, \citenamefont {Messerschmidt}, \citenamefont
  {Williams}, \citenamefont {Koglin}, \citenamefont {Lane}, \citenamefont
  {Hayes}, \citenamefont {Guillet}, \citenamefont {Liang}, \citenamefont
  {Aquila}, \citenamefont {Willmott}, \citenamefont {Robinson}, \citenamefont
  {Gumerlock}, \citenamefont {Botha}, \citenamefont {Nass}, \citenamefont
  {Schlichting}, \citenamefont {Shoeman}, \citenamefont {Stone},\ and\
  \citenamefont {Boutet}}]{stan2016liquid}%
  \BibitemOpen
  \bibfield  {author} {\bibinfo {author} {\bibfnamefont {C.~A.}\ \bibnamefont
  {Stan}}, \bibinfo {author} {\bibfnamefont {D.}~\bibnamefont {Milathianaki}},
  \bibinfo {author} {\bibfnamefont {H.}~\bibnamefont {Laksmono}}, \bibinfo
  {author} {\bibfnamefont {R.~G.}\ \bibnamefont {Sierra}}, \bibinfo {author}
  {\bibfnamefont {T.~A.}\ \bibnamefont {McQueen}}, \bibinfo {author}
  {\bibfnamefont {M.}~\bibnamefont {Messerschmidt}}, \bibinfo {author}
  {\bibfnamefont {G.~J.}\ \bibnamefont {Williams}}, \bibinfo {author}
  {\bibfnamefont {J.~E.}\ \bibnamefont {Koglin}}, \bibinfo {author}
  {\bibfnamefont {T.~J.}\ \bibnamefont {Lane}}, \bibinfo {author}
  {\bibfnamefont {M.~J.}\ \bibnamefont {Hayes}}, \bibinfo {author}
  {\bibfnamefont {S.~A.~H.}\ \bibnamefont {Guillet}}, \bibinfo {author}
  {\bibfnamefont {M.}~\bibnamefont {Liang}}, \bibinfo {author} {\bibfnamefont
  {A.~L.}\ \bibnamefont {Aquila}}, \bibinfo {author} {\bibfnamefont {P.~R.}\
  \bibnamefont {Willmott}}, \bibinfo {author} {\bibfnamefont {J.~S.}\
  \bibnamefont {Robinson}}, \bibinfo {author} {\bibfnamefont {K.~L.}\
  \bibnamefont {Gumerlock}}, \bibinfo {author} {\bibfnamefont {S.}~\bibnamefont
  {Botha}}, \bibinfo {author} {\bibfnamefont {K.}~\bibnamefont {Nass}},
  \bibinfo {author} {\bibfnamefont {I.}~\bibnamefont {Schlichting}}, \bibinfo
  {author} {\bibfnamefont {R.~L.}\ \bibnamefont {Shoeman}}, \bibinfo {author}
  {\bibfnamefont {H.~A.}\ \bibnamefont {Stone}}, \ and\ \bibinfo {author}
  {\bibfnamefont {S.}~\bibnamefont {Boutet}},\ }\href {\doibase
  10.1038/nphys3779} {\bibfield  {journal} {\bibinfo  {journal} {Nat. Phys.}\
  }\textbf {\bibinfo {volume} {12}},\ \bibinfo {pages} {966} (\bibinfo {year}
  {2016}{\natexlab{a}})}\BibitemShut {NoStop}%
\bibitem [{\citenamefont {Stan}\ \emph
  {et~al.}(2016{\natexlab{b}})\citenamefont {Stan}, \citenamefont {Willmott},
  \citenamefont {Stone}, \citenamefont {Koglin}, \citenamefont {Liang},
  \citenamefont {Aquila}, \citenamefont {Robinson}, \citenamefont {Gumerlock},
  \citenamefont {Blaj}, \citenamefont {Sierra}, \citenamefont {Boutet},
  \citenamefont {Guillet}, \citenamefont {Curtis}, \citenamefont {Vetter},
  \citenamefont {Loos}, \citenamefont {Turner},\ and\ \citenamefont
  {Decker}}]{stan2016negative}%
  \BibitemOpen
  \bibfield  {author} {\bibinfo {author} {\bibfnamefont {C.~A.}\ \bibnamefont
  {Stan}}, \bibinfo {author} {\bibfnamefont {P.~R.}\ \bibnamefont {Willmott}},
  \bibinfo {author} {\bibfnamefont {H.~A.}\ \bibnamefont {Stone}}, \bibinfo
  {author} {\bibfnamefont {J.~E.}\ \bibnamefont {Koglin}}, \bibinfo {author}
  {\bibfnamefont {M.}~\bibnamefont {Liang}}, \bibinfo {author} {\bibfnamefont
  {A.~L.}\ \bibnamefont {Aquila}}, \bibinfo {author} {\bibfnamefont {J.~S.}\
  \bibnamefont {Robinson}}, \bibinfo {author} {\bibfnamefont {K.~L.}\
  \bibnamefont {Gumerlock}}, \bibinfo {author} {\bibfnamefont {G.}~\bibnamefont
  {Blaj}}, \bibinfo {author} {\bibfnamefont {R.~G.}\ \bibnamefont {Sierra}},
  \bibinfo {author} {\bibfnamefont {S.}~\bibnamefont {Boutet}}, \bibinfo
  {author} {\bibfnamefont {S.~A.~H.}\ \bibnamefont {Guillet}}, \bibinfo
  {author} {\bibfnamefont {R.~H.}\ \bibnamefont {Curtis}}, \bibinfo {author}
  {\bibfnamefont {S.~L.}\ \bibnamefont {Vetter}}, \bibinfo {author}
  {\bibfnamefont {H.}~\bibnamefont {Loos}}, \bibinfo {author} {\bibfnamefont
  {J.~L.}\ \bibnamefont {Turner}}, \ and\ \bibinfo {author} {\bibfnamefont
  {F.-J.}\ \bibnamefont {Decker}},\ }\href {\doibase
  10.1021/acs.jpclett.6b00687} {\bibfield  {journal} {\bibinfo  {journal} {J.
  Phys. Chem. Lett.}\ }\textbf {\bibinfo {volume} {7}},\ \bibinfo {pages}
  {2055} (\bibinfo {year} {2016}{\natexlab{b}})}\BibitemShut {NoStop}%
\bibitem [{\citenamefont {Vinokhodov}\ \emph {et~al.}(2016)\citenamefont
  {Vinokhodov}, \citenamefont {Koshelev}, \citenamefont {Krivtsun},
  \citenamefont {Krivokorytov}, \citenamefont {Sidelnikov}, \citenamefont
  {Medvedev}, \citenamefont {Kompanets}, \citenamefont {Melnikov},\ and\
  \citenamefont {Chekalin}}]{vinokhodov2016}%
  \BibitemOpen
  \bibfield  {author} {\bibinfo {author} {\bibfnamefont {A.~Y.}\ \bibnamefont
  {Vinokhodov}}, \bibinfo {author} {\bibfnamefont {K.~N.}\ \bibnamefont
  {Koshelev}}, \bibinfo {author} {\bibfnamefont {V.~M.}\ \bibnamefont
  {Krivtsun}}, \bibinfo {author} {\bibfnamefont {M.~S.}\ \bibnamefont
  {Krivokorytov}}, \bibinfo {author} {\bibfnamefont {Y.~V.}\ \bibnamefont
  {Sidelnikov}}, \bibinfo {author} {\bibfnamefont {V.~V.}\ \bibnamefont
  {Medvedev}}, \bibinfo {author} {\bibfnamefont {V.~O.}\ \bibnamefont
  {Kompanets}}, \bibinfo {author} {\bibfnamefont {A.~A.}\ \bibnamefont
  {Melnikov}}, \ and\ \bibinfo {author} {\bibfnamefont {S.~V.}\ \bibnamefont
  {Chekalin}},\ }\href {\doibase 10.1070/QE2016v046n01ABEH015867} {\bibfield
  {journal} {\bibinfo  {journal} {Quantum Electron.}\ }\textbf {\bibinfo
  {volume} {46}},\ \bibinfo {pages} {23} (\bibinfo {year} {2016})}\BibitemShut
  {NoStop}%
\bibitem [{\citenamefont {Basko}\ \emph {et~al.}(2017)\citenamefont {Basko},
  \citenamefont {Krivokorytov}, \citenamefont {Vinokhodov}, \citenamefont
  {Sidelnikov}, \citenamefont {Krivtsun}, \citenamefont {Medvedev},
  \citenamefont {Kim}, \citenamefont {Kompanets}, \citenamefont {Lash},\ and\
  \citenamefont {Koshelev}}]{basko2017}%
  \BibitemOpen
  \bibfield  {author} {\bibinfo {author} {\bibfnamefont {M.~M.}\ \bibnamefont
  {Basko}}, \bibinfo {author} {\bibfnamefont {M.~S.}\ \bibnamefont
  {Krivokorytov}}, \bibinfo {author} {\bibfnamefont {A.~Y.}\ \bibnamefont
  {Vinokhodov}}, \bibinfo {author} {\bibfnamefont {Y.~V.}\ \bibnamefont
  {Sidelnikov}}, \bibinfo {author} {\bibfnamefont {V.~M.}\ \bibnamefont
  {Krivtsun}}, \bibinfo {author} {\bibfnamefont {V.~V.}\ \bibnamefont
  {Medvedev}}, \bibinfo {author} {\bibfnamefont {D.~A.}\ \bibnamefont {Kim}},
  \bibinfo {author} {\bibfnamefont {V.~O.}\ \bibnamefont {Kompanets}}, \bibinfo
  {author} {\bibfnamefont {A.~A.}\ \bibnamefont {Lash}}, \ and\ \bibinfo
  {author} {\bibfnamefont {K.~N.}\ \bibnamefont {Koshelev}},\ }\href {\doibase
  10.1088/1612-202X/aa539b} {\bibfield  {journal} {\bibinfo  {journal} {Laser
  Phys. Lett.}\ }\textbf {\bibinfo {volume} {14}},\ \bibinfo {pages} {036001}
  (\bibinfo {year} {2017})}\BibitemShut {NoStop}%
\bibitem [{\citenamefont {Krivokorytov}\ \emph {et~al.}(2017)\citenamefont
  {Krivokorytov}, \citenamefont {Vinokhodov}, \citenamefont {Sidelnikov},
  \citenamefont {Krivtsun}, \citenamefont {Kompanets}, \citenamefont {Lash},
  \citenamefont {Koshelev},\ and\ \citenamefont {Medvedev}}]{krivokorytov2017}%
  \BibitemOpen
  \bibfield  {author} {\bibinfo {author} {\bibfnamefont {M.~S.}\ \bibnamefont
  {Krivokorytov}}, \bibinfo {author} {\bibfnamefont {A.~Y.}\ \bibnamefont
  {Vinokhodov}}, \bibinfo {author} {\bibfnamefont {Y.~V.}\ \bibnamefont
  {Sidelnikov}}, \bibinfo {author} {\bibfnamefont {V.~M.}\ \bibnamefont
  {Krivtsun}}, \bibinfo {author} {\bibfnamefont {V.~O.}\ \bibnamefont
  {Kompanets}}, \bibinfo {author} {\bibfnamefont {A.~A.}\ \bibnamefont {Lash}},
  \bibinfo {author} {\bibfnamefont {K.~N.}\ \bibnamefont {Koshelev}}, \ and\
  \bibinfo {author} {\bibfnamefont {V.~V.}\ \bibnamefont {Medvedev}},\ }\href
  {\doibase 10.1103/PhysRevE.95.031101} {\bibfield  {journal} {\bibinfo
  {journal} {Phys. Rev. E}\ }\textbf {\bibinfo {volume} {95}},\ \bibinfo
  {pages} {031101} (\bibinfo {year} {2017})}\BibitemShut {NoStop}%
\bibitem [{\citenamefont {Banine}\ \emph {et~al.}(2011)\citenamefont {Banine},
  \citenamefont {Koshelev},\ and\ \citenamefont {Swinkels}}]{Banine2011}%
  \BibitemOpen
  \bibfield  {author} {\bibinfo {author} {\bibfnamefont {V.~Y.}\ \bibnamefont
  {Banine}}, \bibinfo {author} {\bibfnamefont {K.~N.}\ \bibnamefont
  {Koshelev}}, \ and\ \bibinfo {author} {\bibfnamefont {G.~H. P.~M.}\
  \bibnamefont {Swinkels}},\ }\href {\doibase 10.1088/0022-3727/44/25/253001}
  {\bibfield  {journal} {\bibinfo  {journal} {J. Phys. D: Appl. Phys.}\
  }\textbf {\bibinfo {volume} {44}},\ \bibinfo {pages} {253001} (\bibinfo
  {year} {2011})}\BibitemShut {NoStop}%
\bibitem [{\citenamefont {Pirati}\ \emph {et~al.}(2017)\citenamefont {Pirati},
  \citenamefont {van Schoot}, \citenamefont {Troost}, \citenamefont {van
  Ballegoij}, \citenamefont {Krabbendam}, \citenamefont {Stoeldraijer},
  \citenamefont {Loopstra}, \citenamefont {Benschop}, \citenamefont {Finders},
  \citenamefont {Meiling}, \citenamefont {van Setten}, \citenamefont {Mika},
  \citenamefont {Dredonx}, \citenamefont {Stamm}, \citenamefont {Kneer},
  \citenamefont {Thuering}, \citenamefont {Kaiser}, \citenamefont {Heil},\ and\
  \citenamefont {Migura}}]{Pirati2017}%
  \BibitemOpen
  \bibfield  {author} {\bibinfo {author} {\bibfnamefont {A.}~\bibnamefont
  {Pirati}}, \bibinfo {author} {\bibfnamefont {J.}~\bibnamefont {van Schoot}},
  \bibinfo {author} {\bibfnamefont {K.}~\bibnamefont {Troost}}, \bibinfo
  {author} {\bibfnamefont {R.}~\bibnamefont {van Ballegoij}}, \bibinfo {author}
  {\bibfnamefont {P.}~\bibnamefont {Krabbendam}}, \bibinfo {author}
  {\bibfnamefont {J.}~\bibnamefont {Stoeldraijer}}, \bibinfo {author}
  {\bibfnamefont {E.}~\bibnamefont {Loopstra}}, \bibinfo {author}
  {\bibfnamefont {J.}~\bibnamefont {Benschop}}, \bibinfo {author}
  {\bibfnamefont {J.}~\bibnamefont {Finders}}, \bibinfo {author} {\bibfnamefont
  {H.}~\bibnamefont {Meiling}}, \bibinfo {author} {\bibfnamefont
  {E.}~\bibnamefont {van Setten}}, \bibinfo {author} {\bibfnamefont
  {N.}~\bibnamefont {Mika}}, \bibinfo {author} {\bibfnamefont {J.}~\bibnamefont
  {Dredonx}}, \bibinfo {author} {\bibfnamefont {U.}~\bibnamefont {Stamm}},
  \bibinfo {author} {\bibfnamefont {B.}~\bibnamefont {Kneer}}, \bibinfo
  {author} {\bibfnamefont {B.}~\bibnamefont {Thuering}}, \bibinfo {author}
  {\bibfnamefont {W.}~\bibnamefont {Kaiser}}, \bibinfo {author} {\bibfnamefont
  {T.}~\bibnamefont {Heil}}, \ and\ \bibinfo {author} {\bibfnamefont
  {S.}~\bibnamefont {Migura}},\ }\href {\doibase 10.1117/12.2261079} {\bibfield
   {journal} {\bibinfo  {journal} {Proc. SPIE}\ }\textbf {\bibinfo {volume}
  {10143}},\ \bibinfo {pages} {10143 } (\bibinfo {year} {2017})}\BibitemShut
  {NoStop}%
\bibitem [{\citenamefont {Fomenkov}\ \emph {et~al.}(2017)\citenamefont
  {Fomenkov}, \citenamefont {Brandt}, \citenamefont {Ershov}, \citenamefont
  {Schafgans}, \citenamefont {Tao}, \citenamefont {Vaschenko}, \citenamefont
  {Rokitski}, \citenamefont {Kats}, \citenamefont {Vargas}, \citenamefont
  {Purvis}, \citenamefont {Rafac}, \citenamefont {{La Fontaine}}, \citenamefont
  {{De Dea}}, \citenamefont {LaForge}, \citenamefont {Stewart}, \citenamefont
  {Chang}, \citenamefont {Graham}, \citenamefont {Riggs}, \citenamefont
  {Taylor}, \citenamefont {Abraham},\ and\ \citenamefont
  {Brown}}]{Fomenkov2017}%
  \BibitemOpen
  \bibfield  {author} {\bibinfo {author} {\bibfnamefont {I.}~\bibnamefont
  {Fomenkov}}, \bibinfo {author} {\bibfnamefont {D.}~\bibnamefont {Brandt}},
  \bibinfo {author} {\bibfnamefont {A.}~\bibnamefont {Ershov}}, \bibinfo
  {author} {\bibfnamefont {A.}~\bibnamefont {Schafgans}}, \bibinfo {author}
  {\bibfnamefont {Y.}~\bibnamefont {Tao}}, \bibinfo {author} {\bibfnamefont
  {G.}~\bibnamefont {Vaschenko}}, \bibinfo {author} {\bibfnamefont
  {S.}~\bibnamefont {Rokitski}}, \bibinfo {author} {\bibfnamefont
  {M.}~\bibnamefont {Kats}}, \bibinfo {author} {\bibfnamefont {M.}~\bibnamefont
  {Vargas}}, \bibinfo {author} {\bibfnamefont {M.}~\bibnamefont {Purvis}},
  \bibinfo {author} {\bibfnamefont {R.}~\bibnamefont {Rafac}}, \bibinfo
  {author} {\bibfnamefont {B.}~\bibnamefont {{La Fontaine}}}, \bibinfo {author}
  {\bibfnamefont {S.}~\bibnamefont {{De Dea}}}, \bibinfo {author}
  {\bibfnamefont {A.}~\bibnamefont {LaForge}}, \bibinfo {author} {\bibfnamefont
  {J.}~\bibnamefont {Stewart}}, \bibinfo {author} {\bibfnamefont
  {S.}~\bibnamefont {Chang}}, \bibinfo {author} {\bibfnamefont
  {M.}~\bibnamefont {Graham}}, \bibinfo {author} {\bibfnamefont
  {D.}~\bibnamefont {Riggs}}, \bibinfo {author} {\bibfnamefont
  {T.}~\bibnamefont {Taylor}}, \bibinfo {author} {\bibfnamefont
  {M.}~\bibnamefont {Abraham}}, \ and\ \bibinfo {author} {\bibfnamefont
  {D.}~\bibnamefont {Brown}},\ }\href {\doibase 10.1515/aot-2017-0029}
  {\bibfield  {journal} {\bibinfo  {journal} {Adv. Opt. Technol.}\ }\textbf
  {\bibinfo {volume} {6}},\ \bibinfo {pages} {173} (\bibinfo {year}
  {2017})}\BibitemShut {NoStop}%
\bibitem [{\citenamefont {Kawasuji}\ \emph {et~al.}(2017)\citenamefont
  {Kawasuji}, \citenamefont {Nowak}, \citenamefont {Hori}, \citenamefont
  {Okamoto}, \citenamefont {Tanaka}, \citenamefont {Watanabe}, \citenamefont
  {Abe}, \citenamefont {Kodama}, \citenamefont {Nakarai}, \citenamefont
  {Yamazaki}, \citenamefont {Okazaki}, \citenamefont {Saitou}, \citenamefont
  {Mizoguchi},\ and\ \citenamefont {Shiraishi}}]{Kawasuji2017}%
  \BibitemOpen
  \bibfield  {author} {\bibinfo {author} {\bibfnamefont {Y.}~\bibnamefont
  {Kawasuji}}, \bibinfo {author} {\bibfnamefont {K.~M.}\ \bibnamefont {Nowak}},
  \bibinfo {author} {\bibfnamefont {T.}~\bibnamefont {Hori}}, \bibinfo {author}
  {\bibfnamefont {T.}~\bibnamefont {Okamoto}}, \bibinfo {author} {\bibfnamefont
  {H.}~\bibnamefont {Tanaka}}, \bibinfo {author} {\bibfnamefont
  {Y.}~\bibnamefont {Watanabe}}, \bibinfo {author} {\bibfnamefont
  {T.}~\bibnamefont {Abe}}, \bibinfo {author} {\bibfnamefont {T.}~\bibnamefont
  {Kodama}}, \bibinfo {author} {\bibfnamefont {H.}~\bibnamefont {Nakarai}},
  \bibinfo {author} {\bibfnamefont {T.}~\bibnamefont {Yamazaki}}, \bibinfo
  {author} {\bibfnamefont {S.}~\bibnamefont {Okazaki}}, \bibinfo {author}
  {\bibfnamefont {T.}~\bibnamefont {Saitou}}, \bibinfo {author} {\bibfnamefont
  {H.}~\bibnamefont {Mizoguchi}}, \ and\ \bibinfo {author} {\bibfnamefont
  {Y.}~\bibnamefont {Shiraishi}},\ }\href {\doibase 10.1117/12.2257808}
  {\bibfield  {journal} {\bibinfo  {journal} {Proc. SPIE}\ }\textbf {\bibinfo
  {volume} {10143}},\ \bibinfo {pages} {10143 } (\bibinfo {year}
  {2017})}\BibitemShut {NoStop}%
\bibitem [{\citenamefont {Krivokorytov}\ \emph {et~al.}(2018)\citenamefont
  {Krivokorytov}, \citenamefont {Zeng}, \citenamefont {Lakatosh}, \citenamefont
  {Vinokhodov}, \citenamefont {Sidelnikov}, \citenamefont {Kompanets},
  \citenamefont {Krivtsun}, \citenamefont {Koshelev}, \citenamefont {Ohl},\
  and\ \citenamefont {Medvedev}}]{Kriv2018}%
  \BibitemOpen
  \bibfield  {author} {\bibinfo {author} {\bibfnamefont {M.~S.}\ \bibnamefont
  {Krivokorytov}}, \bibinfo {author} {\bibfnamefont {Q.}~\bibnamefont {Zeng}},
  \bibinfo {author} {\bibfnamefont {B.~V.}\ \bibnamefont {Lakatosh}}, \bibinfo
  {author} {\bibfnamefont {A.~Y.}\ \bibnamefont {Vinokhodov}}, \bibinfo
  {author} {\bibfnamefont {Y.~V.}\ \bibnamefont {Sidelnikov}}, \bibinfo
  {author} {\bibfnamefont {V.~O.}\ \bibnamefont {Kompanets}}, \bibinfo {author}
  {\bibfnamefont {V.~M.}\ \bibnamefont {Krivtsun}}, \bibinfo {author}
  {\bibfnamefont {K.~N.}\ \bibnamefont {Koshelev}}, \bibinfo {author}
  {\bibfnamefont {C.-D.}\ \bibnamefont {Ohl}}, \ and\ \bibinfo {author}
  {\bibfnamefont {V.~V.}\ \bibnamefont {Medvedev}},\ }\href {\doibase
  10.1038/s41598-017-19140-w} {\bibfield  {journal} {\bibinfo  {journal} {Sci.
  Rep.}\ }\textbf {\bibinfo {volume} {8}},\ \bibinfo {pages} {597} (\bibinfo
  {year} {2018})}\BibitemShut {NoStop}%
\bibitem [{\citenamefont {Kurilovich}\ \emph {et~al.}(2016)\citenamefont
  {Kurilovich}, \citenamefont {Klein}, \citenamefont {Torretti}, \citenamefont
  {Lassise}, \citenamefont {Hoekstra}, \citenamefont {Ubachs}, \citenamefont
  {Gelderblom},\ and\ \citenamefont {Versolato}}]{Kurilovich2016}%
  \BibitemOpen
  \bibfield  {author} {\bibinfo {author} {\bibfnamefont {D.}~\bibnamefont
  {Kurilovich}}, \bibinfo {author} {\bibfnamefont {A.~L.}\ \bibnamefont
  {Klein}}, \bibinfo {author} {\bibfnamefont {F.}~\bibnamefont {Torretti}},
  \bibinfo {author} {\bibfnamefont {A.}~\bibnamefont {Lassise}}, \bibinfo
  {author} {\bibfnamefont {R.}~\bibnamefont {Hoekstra}}, \bibinfo {author}
  {\bibfnamefont {W.}~\bibnamefont {Ubachs}}, \bibinfo {author} {\bibfnamefont
  {H.}~\bibnamefont {Gelderblom}}, \ and\ \bibinfo {author} {\bibfnamefont
  {O.~O.}\ \bibnamefont {Versolato}},\ }\href {\doibase
  10.1103/PhysRevApplied.6.014018} {\bibfield  {journal} {\bibinfo  {journal}
  {Phys. Rev. Appl.}\ }\textbf {\bibinfo {volume} {6}},\ \bibinfo {pages}
  {014018} (\bibinfo {year} {2016})}\BibitemShut {NoStop}%
\bibitem [{\citenamefont {{Stodolna}}\ \emph {et~al.}(2018)\citenamefont
  {{Stodolna}}, \citenamefont {{de Faria Pinto}}, \citenamefont {{Ali}},
  \citenamefont {{Bayerle}}, \citenamefont {{Kurilovich}}, \citenamefont
  {{Mathijssen}}, \citenamefont {{Hoekstra}}, \citenamefont {{Versolato}},
  \citenamefont {{Eikema}},\ and\ \citenamefont {{Witte}}}]{Stodolna2018}%
  \BibitemOpen
  \bibfield  {author} {\bibinfo {author} {\bibfnamefont {A.~S.}\ \bibnamefont
  {{Stodolna}}}, \bibinfo {author} {\bibfnamefont {T.}~\bibnamefont {{de Faria
  Pinto}}}, \bibinfo {author} {\bibfnamefont {F.}~\bibnamefont {{Ali}}},
  \bibinfo {author} {\bibfnamefont {A.}~\bibnamefont {{Bayerle}}}, \bibinfo
  {author} {\bibfnamefont {D.}~\bibnamefont {{Kurilovich}}}, \bibinfo {author}
  {\bibfnamefont {J.}~\bibnamefont {{Mathijssen}}}, \bibinfo {author}
  {\bibfnamefont {R.}~\bibnamefont {{Hoekstra}}}, \bibinfo {author}
  {\bibfnamefont {O.~O.}\ \bibnamefont {{Versolato}}}, \bibinfo {author}
  {\bibfnamefont {K.~S.~E.}\ \bibnamefont {{Eikema}}}, \ and\ \bibinfo {author}
  {\bibfnamefont {S.}~\bibnamefont {{Witte}}},\ }\href@noop {} {\bibfield
  {journal} {\bibinfo  {journal} {ArXiv e-prints}\ } (\bibinfo {year}
  {2018})},\ \Eprint {http://arxiv.org/abs/1804.01329} {arXiv:1804.01329
  [physics.plasm-ph]} \BibitemShut {NoStop}%
\bibitem [{SM()}]{SM}%
  \BibitemOpen
  \href@noop {} {}\bibinfo {note} {See Supplemental Material at [URL will be
  inserted by publisher] for stroboscopic shadowgraphy movies}\BibitemShut
  {NoStop}%
\bibitem [{\citenamefont {Sokolowski-Tinten}\ \emph {et~al.}(1998)\citenamefont
  {Sokolowski-Tinten}, \citenamefont {Bialkowski}, \citenamefont {Cavalleri},
  \citenamefont {von~der Linde}, \citenamefont {Oparin}, \citenamefont
  {{Meyer-ter-Vehn}},\ and\ \citenamefont {Anisimov}}]{Tinten1998}%
  \BibitemOpen
  \bibfield  {author} {\bibinfo {author} {\bibfnamefont {K.}~\bibnamefont
  {Sokolowski-Tinten}}, \bibinfo {author} {\bibfnamefont {J.}~\bibnamefont
  {Bialkowski}}, \bibinfo {author} {\bibfnamefont {A.}~\bibnamefont
  {Cavalleri}}, \bibinfo {author} {\bibfnamefont {D.}~\bibnamefont {von~der
  Linde}}, \bibinfo {author} {\bibfnamefont {A.}~\bibnamefont {Oparin}},
  \bibinfo {author} {\bibfnamefont {J.}~\bibnamefont {{Meyer-ter-Vehn}}}, \
  and\ \bibinfo {author} {\bibfnamefont {S.~I.}\ \bibnamefont {Anisimov}},\
  }\href {\doibase 10.1103/PhysRevLett.81.224} {\bibfield  {journal} {\bibinfo
  {journal} {Phys. Rev. Lett.}\ }\textbf {\bibinfo {volume} {81}},\ \bibinfo
  {pages} {224} (\bibinfo {year} {1998})}\BibitemShut {NoStop}%
\bibitem [{\citenamefont {von~der Linde}\ and\ \citenamefont
  {Sokolowski-Tinten}(2000)}]{VONDERLINDE2000}%
  \BibitemOpen
  \bibfield  {author} {\bibinfo {author} {\bibfnamefont {D.}~\bibnamefont
  {von~der Linde}}\ and\ \bibinfo {author} {\bibfnamefont {K.}~\bibnamefont
  {Sokolowski-Tinten}},\ }\href {\doibase
  https://doi.org/10.1016/S0169-4332(99)00440-7} {\bibfield  {journal}
  {\bibinfo  {journal} {Appl. Surf. Sci.}\ }\textbf {\bibinfo {volume}
  {154-155}},\ \bibinfo {pages} {1 } (\bibinfo {year} {2000})}\BibitemShut
  {NoStop}%
\bibitem [{\citenamefont {Reijers}\ \emph {et~al.}(2017)\citenamefont
  {Reijers}, \citenamefont {Snoeijer},\ and\ \citenamefont
  {Gelderblom}}]{reijers2017}%
  \BibitemOpen
  \bibfield  {author} {\bibinfo {author} {\bibfnamefont {S.~A.}\ \bibnamefont
  {Reijers}}, \bibinfo {author} {\bibfnamefont {J.~H.}\ \bibnamefont
  {Snoeijer}}, \ and\ \bibinfo {author} {\bibfnamefont {H.}~\bibnamefont
  {Gelderblom}},\ }\href {\doibase 10.1017/jfm.2017.518} {\bibfield  {journal}
  {\bibinfo  {journal} {J. Fluid Mech.}\ }\textbf {\bibinfo {volume} {828}},\
  \bibinfo {pages} {374} (\bibinfo {year} {2017})}\BibitemShut {NoStop}%
\bibitem [{\citenamefont {Thoroddsen}\ \emph {et~al.}(2009)\citenamefont
  {Thoroddsen}, \citenamefont {Takehara}, \citenamefont {Etoh},\ and\
  \citenamefont {Ohl}}]{thoroddsen2009}%
  \BibitemOpen
  \bibfield  {author} {\bibinfo {author} {\bibfnamefont {S.~T.}\ \bibnamefont
  {Thoroddsen}}, \bibinfo {author} {\bibfnamefont {K.}~\bibnamefont
  {Takehara}}, \bibinfo {author} {\bibfnamefont {T.}~\bibnamefont {Etoh}}, \
  and\ \bibinfo {author} {\bibfnamefont {C.-D.}\ \bibnamefont {Ohl}},\ }\href
  {\doibase 10.1063/1.3253394} {\bibfield  {journal} {\bibinfo  {journal}
  {Phys. Fluids}\ }\textbf {\bibinfo {volume} {21}},\ \bibinfo {pages} {112101}
  (\bibinfo {year} {2009})}\BibitemShut {NoStop}%
\bibitem [{\citenamefont {Gonzalez-Avila}\ \emph {et~al.}(2011)\citenamefont
  {Gonzalez-Avila}, \citenamefont {Klaseboer}, \citenamefont {Khoo},\ and\
  \citenamefont {Ohl}}]{gonzalez2011}%
  \BibitemOpen
  \bibfield  {author} {\bibinfo {author} {\bibfnamefont {S.~R.}\ \bibnamefont
  {Gonzalez-Avila}}, \bibinfo {author} {\bibfnamefont {E.}~\bibnamefont
  {Klaseboer}}, \bibinfo {author} {\bibfnamefont {B.~C.}\ \bibnamefont {Khoo}},
  \ and\ \bibinfo {author} {\bibfnamefont {C.-D.}\ \bibnamefont {Ohl}},\ }\href
  {\doibase 10.1017/jfm.2011.212} {\bibfield  {journal} {\bibinfo  {journal}
  {J. Fluid Mech.}\ }\textbf {\bibinfo {volume} {682}},\ \bibinfo {pages} {241}
  (\bibinfo {year} {2011})}\BibitemShut {NoStop}%
\bibitem [{\citenamefont {Eidmann}\ \emph {et~al.}(1984)\citenamefont
  {Eidmann}, \citenamefont {Amiranoff}, \citenamefont {Fedosejevs},
  \citenamefont {Maaswinkel}, \citenamefont {Petsch}, \citenamefont {Sigel},
  \citenamefont {Spindler}, \citenamefont {Teng}, \citenamefont {Tsakiris},\
  and\ \citenamefont {Witkowski}}]{Eidmann1984}%
  \BibitemOpen
  \bibfield  {author} {\bibinfo {author} {\bibfnamefont {K.}~\bibnamefont
  {Eidmann}}, \bibinfo {author} {\bibfnamefont {F.}~\bibnamefont {Amiranoff}},
  \bibinfo {author} {\bibfnamefont {R.}~\bibnamefont {Fedosejevs}}, \bibinfo
  {author} {\bibfnamefont {A.~G.~M.}\ \bibnamefont {Maaswinkel}}, \bibinfo
  {author} {\bibfnamefont {R.}~\bibnamefont {Petsch}}, \bibinfo {author}
  {\bibfnamefont {R.}~\bibnamefont {Sigel}}, \bibinfo {author} {\bibfnamefont
  {G.}~\bibnamefont {Spindler}}, \bibinfo {author} {\bibfnamefont {Y.-l.}\
  \bibnamefont {Teng}}, \bibinfo {author} {\bibfnamefont {G.}~\bibnamefont
  {Tsakiris}}, \ and\ \bibinfo {author} {\bibfnamefont {S.}~\bibnamefont
  {Witkowski}},\ }\href {\doibase 10.1103/PhysRevA.30.2568} {\bibfield
  {journal} {\bibinfo  {journal} {Phys. Rev. A}\ }\textbf {\bibinfo {volume}
  {30}},\ \bibinfo {pages} {2568} (\bibinfo {year} {1984})}\BibitemShut
  {NoStop}%
\bibitem [{\citenamefont {Dahmani}(1993)}]{Dahmani1993}%
  \BibitemOpen
  \bibfield  {author} {\bibinfo {author} {\bibfnamefont {F.}~\bibnamefont
  {Dahmani}},\ }\href {\doibase 10.1063/1.355276} {\bibfield  {journal}
  {\bibinfo  {journal} {J. Appl. Phys.}\ }\textbf {\bibinfo {volume} {74}},\
  \bibinfo {pages} {622} (\bibinfo {year} {1993})}\BibitemShut {NoStop}%
\bibitem [{\citenamefont {Eidmann}\ \emph {et~al.}(2000)\citenamefont
  {Eidmann}, \citenamefont {{Meyer-ter-Vehn}}, \citenamefont {Schlegel},\ and\
  \citenamefont {H\"uller}}]{Eidmann2000}%
  \BibitemOpen
  \bibfield  {author} {\bibinfo {author} {\bibfnamefont {K.}~\bibnamefont
  {Eidmann}}, \bibinfo {author} {\bibfnamefont {J.}~\bibnamefont
  {{Meyer-ter-Vehn}}}, \bibinfo {author} {\bibfnamefont {T.}~\bibnamefont
  {Schlegel}}, \ and\ \bibinfo {author} {\bibfnamefont {S.}~\bibnamefont
  {H\"uller}},\ }\href {\doibase 10.1103/PhysRevE.62.1202} {\bibfield
  {journal} {\bibinfo  {journal} {Phys. Rev. E}\ }\textbf {\bibinfo {volume}
  {62}},\ \bibinfo {pages} {1202} (\bibinfo {year} {2000})}\BibitemShut
  {NoStop}%
\bibitem [{\citenamefont {Vledouts}\ \emph {et~al.}(2016)\citenamefont
  {Vledouts}, \citenamefont {Quinard}, \citenamefont {Vandenberghe},\ and\
  \citenamefont {Villermaux}}]{Vledouts2016}%
  \BibitemOpen
  \bibfield  {author} {\bibinfo {author} {\bibfnamefont {A.}~\bibnamefont
  {Vledouts}}, \bibinfo {author} {\bibfnamefont {J.}~\bibnamefont {Quinard}},
  \bibinfo {author} {\bibfnamefont {N.}~\bibnamefont {Vandenberghe}}, \ and\
  \bibinfo {author} {\bibfnamefont {E.}~\bibnamefont {Villermaux}},\ }\href
  {\doibase 10.1017/jfm.2015.716} {\bibfield  {journal} {\bibinfo  {journal}
  {J. Fluid Mech.}\ }\textbf {\bibinfo {volume} {788}},\ \bibinfo {pages} {246}
  (\bibinfo {year} {2016})}\BibitemShut {NoStop}%
\bibitem [{\citenamefont {{Gelderblom}}\ \emph {et~al.}(2016)\citenamefont
  {{Gelderblom}}, \citenamefont {{Lhuissier}}, \citenamefont {{Klein}},
  \citenamefont {{Bouwhuis}}, \citenamefont {{Lohse}}, \citenamefont
  {{Villermaux}},\ and\ \citenamefont {{Snoeijer}}}]{Gelderblom:2015}%
  \BibitemOpen
  \bibfield  {author} {\bibinfo {author} {\bibfnamefont {H.}~\bibnamefont
  {{Gelderblom}}}, \bibinfo {author} {\bibfnamefont {H.}~\bibnamefont
  {{Lhuissier}}}, \bibinfo {author} {\bibfnamefont {A.~L.}\ \bibnamefont
  {{Klein}}}, \bibinfo {author} {\bibfnamefont {W.}~\bibnamefont {{Bouwhuis}}},
  \bibinfo {author} {\bibfnamefont {D.}~\bibnamefont {{Lohse}}}, \bibinfo
  {author} {\bibfnamefont {E.}~\bibnamefont {{Villermaux}}}, \ and\ \bibinfo
  {author} {\bibfnamefont {J.~H.}\ \bibnamefont {{Snoeijer}}},\ }\href
  {\doibase 10.1017/jfm.2016.182} {\bibfield  {journal} {\bibinfo  {journal}
  {J. Fluid Mech.}\ }\textbf {\bibinfo {volume} {794}},\ \bibinfo {pages} {676}
  (\bibinfo {year} {2016})}\BibitemShut {NoStop}%
\bibitem [{\citenamefont {Keller}\ and\ \citenamefont
  {Kolodner}(1954)}]{Keller1954}%
  \BibitemOpen
  \bibfield  {author} {\bibinfo {author} {\bibfnamefont {J.~B.}\ \bibnamefont
  {Keller}}\ and\ \bibinfo {author} {\bibfnamefont {I.}~\bibnamefont
  {Kolodner}},\ }\href {\doibase 10.1063/1.1721770} {\bibfield  {journal}
  {\bibinfo  {journal} {J. Appl. Phys.}\ }\textbf {\bibinfo {volume} {25}},\
  \bibinfo {pages} {918} (\bibinfo {year} {1954})}\BibitemShut {NoStop}%
\bibitem [{\citenamefont {Chandrasekhar}(1961)}]{Chandrasekhar1961}%
  \BibitemOpen
  \bibfield  {author} {\bibinfo {author} {\bibfnamefont {S.}~\bibnamefont
  {Chandrasekhar}},\ }\href@noop {} {\emph {\bibinfo {title} {Hydrodynamic and
  Hydromagnetic Stability}}}\ (\bibinfo  {publisher} {Oxford University
  Press},\ \bibinfo {year} {1961})\BibitemShut {NoStop}%
\bibitem [{\citenamefont {Bremond}\ and\ \citenamefont
  {Villermaux}(2005)}]{bremond2005bursting}%
  \BibitemOpen
  \bibfield  {author} {\bibinfo {author} {\bibfnamefont {N.}~\bibnamefont
  {Bremond}}\ and\ \bibinfo {author} {\bibfnamefont {E.}~\bibnamefont
  {Villermaux}},\ }\href {\doibase 10.1017/S0022112004002411} {\bibfield
  {journal} {\bibinfo  {journal} {J. Fluid Mech.}\ }\textbf {\bibinfo {volume}
  {524}},\ \bibinfo {pages} {121} (\bibinfo {year} {2005})}\BibitemShut
  {NoStop}%
\bibitem [{\citenamefont {Klein}(2017)}]{KleinPhDThesis}%
  \BibitemOpen
  \bibfield  {author} {\bibinfo {author} {\bibfnamefont {A.}~\bibnamefont
  {Klein}},\ }\emph {\bibinfo {title} {Laser impact on flying drops}},\ \href
  {\doibase 10.3990/1.9789036543422} {Ph.D. thesis},\ \bibinfo  {school}
  {University of Twente} (\bibinfo {year} {2017})\BibitemShut {NoStop}%
\bibitem [{\citenamefont {Klein}\ \emph {et~al.}(2015)\citenamefont {Klein},
  \citenamefont {Bouwhuis}, \citenamefont {Visser}, \citenamefont {Lhuissier},
  \citenamefont {Sun}, \citenamefont {Snoeijer}, \citenamefont {Villermaux},
  \citenamefont {Lohse},\ and\ \citenamefont {Gelderblom}}]{Klein2015}%
  \BibitemOpen
  \bibfield  {author} {\bibinfo {author} {\bibfnamefont {A.~L.}\ \bibnamefont
  {Klein}}, \bibinfo {author} {\bibfnamefont {W.}~\bibnamefont {Bouwhuis}},
  \bibinfo {author} {\bibfnamefont {C.~W.}\ \bibnamefont {Visser}}, \bibinfo
  {author} {\bibfnamefont {H.}~\bibnamefont {Lhuissier}}, \bibinfo {author}
  {\bibfnamefont {C.}~\bibnamefont {Sun}}, \bibinfo {author} {\bibfnamefont
  {J.~H.}\ \bibnamefont {Snoeijer}}, \bibinfo {author} {\bibfnamefont
  {E.}~\bibnamefont {Villermaux}}, \bibinfo {author} {\bibfnamefont
  {D.}~\bibnamefont {Lohse}}, \ and\ \bibinfo {author} {\bibfnamefont
  {H.}~\bibnamefont {Gelderblom}},\ }\href {\doibase
  10.1103/PhysRevApplied.3.044018} {\bibfield  {journal} {\bibinfo  {journal}
  {Phys. Rev. Appl.}\ }\textbf {\bibinfo {volume} {3}},\ \bibinfo {pages}
  {044018} (\bibinfo {year} {2015})}\BibitemShut {NoStop}%
\end{thebibliography}%
\end{document}